\documentclass[fleqn,10pt]{article}
\usepackage{epsfig}
\newcommand{\be}{\begin{equation}}
\newcommand{\ee}{\end{equation}}

\def\lpa{\lambda_{p{-}\rm air}}
\def\spa{\sigma_{p{-}\rm air}}
\def\spai{\sigma_{p{-}\rm air}^{\rm inel}}
\def\spae{\sigma_{p{-}\rm air}^{\rm el}}
\def\spaqe{\sigma_{p{-}\rm air}^{q{-}\rm el}}
\begin{document}
\setcounter{secnumdepth}{4}
\renewcommand\thepage{\ }
%
% Start title input here.
%
\begin{titlepage} % This environment allows me to set up an "empty"
          % page. I must fill it with \title, \abstract and \date.
     % Make the NU logo here.
     % We make a box of the perfect size for the current report number.
%
     % We must ENTER Report Number and DATE below.
\newcommand\reportnumber{706} % ENTER current report number.
\newcommand\mydate{April 2000} %Set date up in logo.
\newlength{\nulogo} % define a new length "nulogo".
     % Use  "\settowidth" to get correct width of box, called "nulogo".
\settowidth{\nulogo}{\small\sf{Northwestern University:}}
\title{
\vspace{-.8in} % Puts everything toward top of page
\hfill\fbox{{\parbox{\nulogo}{\small\sf{Northwestern University:\\
NUHEP \reportnumber\\[.3ex]
University of Wisconsin:\\
MADPH-00-1172\\[.3ex]
\mydate}}}}
\vspace{0.5in} \\
          % The "\fbox" puts a BOX around text and the
          % "\small\sf" sets it in small, sans serif type,
          % the "\parbox{nulogo}{} makes a paragraph box, of correct width,
          % the "\fbox makes a box around the "parbox", and finally,
          % the "\hfill" pushes it to right.
          % Finally, the "\vspace{1in}" sets the NEXT line down by one inch.
% We now continue to put in MAIN title.
{
Extending the Frontiers---Reconciling Accelerator and Cosmic Ray p-p Cross Sections
}}

\author{
M.~M.~Block
\thanks{Work partially supported by Department of Energy contract
DA-AC02-76-Er02289 Task B.}\vspace{-5pt}   \\
{\small\em Department of Physics and Astronomy,} \vspace{-5pt} \\ %Make
    %smaller separation between lines.
{\small\em Northwestern University, Evanston, IL 60208}\\
\vspace{-5pt}\\
   % The negative \vspace decreases the space between last line.
%
F.~Halzen
\thanks{Work partially supported by
Department of Energy Grant No.~DE-FG02-95ER40896 and
the University of Wisconsin Research
Committee with funds granted by the Wisconsin Alumni Research Foundation.}
\vspace{-5pt} \\
   % The negative \vspace decreasesthe space between last line.
{\small\em Department of Physics,} \vspace{-5pt} \\
{\small\em University of Wisconsin, Madison, WI 53706}  \\
\vspace{-5pt}\\
   % The negative \vspace decreasesthe space between last line.
%
T. Stanev
\thanks{Work partially supported by the U.S.~Department of Energy under Grant  
No. DE-FG02-91ER40626.}
\vspace{-5pt} \\
{\small\em Bartol Research Institute, University of Delaware, Newark, DE 19716
}\vspace{-5pt}  \\
\vspace{-5pt}\\
   % The negative \vspace decreasesthe space between last line.
%
}    %    End of title section.
\vfill
\vspace{.5in}
     %    End of title section.
     % Command to print title.
%\date{} % This stops printing of date.
\date {}%{April 23, 2000}
\maketitle
\begin{abstract}
%\vskip-5ex %% reduce space between "Abstract" and it's text
We simultaneously fit a QCD-inspired parameterization
 of all accelerator data on forward proton-proton and antiproton-proton
 scattering amplitudes, {\em together} with cosmic ray data (using Glauber  
theory), to predict proton-air
 and proton-proton cross sections at energies near $\sqrt s \approx$ 30 TeV.   
The p-air cosmic ray
 measurements provide a strong constraint on the inclusive
 particle production cross section, as well as greatly reducing the errors on  
the fit parameters---in turn, greatly reducing the errors in the high energy  
proton-proton and proton-air cross section predictions.
\end{abstract}
\end{titlepage} % End of titlepage environment.
%
% Start MAIN body of text here.
     % Turn on page numbering, in arabic numerals..
\pagenumbering{arabic}
\renewcommand{\thepage}{-- \arabic{page}\ --}  % Put dashes on either
          % side of the page number.
%    Text starts here.
%

The energy range of cosmic ray  experiments covers not only the energy of the
Large Hadron Collider (LHC), but extends beyond it. Cosmic ray experiments can  
measure the penetration in the atmosphere
 of these very high energy protons---however, extracting
 proton-proton cross sections from cosmic ray observations is far from
 straightforward~\cite{gaisser}. By a variety of experimental techniques,
 cosmic ray experiments map the atmospheric depth at which cosmic ray
 initiated showers develop.
 The measured quantity is the shower attenuation length ($\Lambda_m$), which  
is not only
 sensitive to the interaction length of the protons in the atmosphere
 ($\lpa$), with
\begin{equation}
\Lambda_m = k \lpa = k { 14.5 m_p \over \spai} \,,  \label{eq:Lambda_m}
\end{equation}
 but also depends critically on the inelasticity, which determines the rate at  
which the energy of the primary proton
 is dissipated into electromagnetic shower energy observed in the
 experiment. The latter effect is taken into account in Eq.\,(\ref{eq:Lambda_m})
 by the parameter $k$; $m_p$ is the proton mass and $\spai$ the inelastic
 proton-air cross section. The departure of $k$ from unity depends on the inclusive
 particle production cross section in nucleon and meson interactions
 on the light nuclear target of the atmosphere and its energy dependence.

\begin{figure}[h]%Fig. 1
\begin{center}
\mbox{\epsfig{file=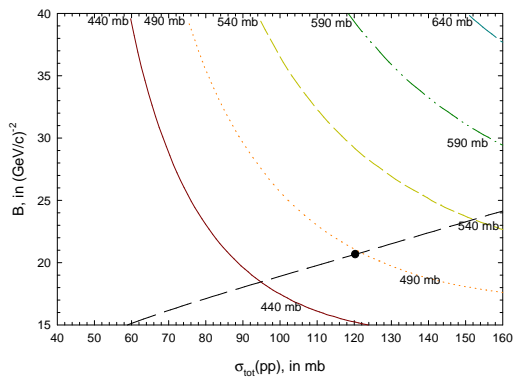%
              ,width=3in,bbllx=55pt,bblly=220pt,bburx=520pt,bbury=560pt,clip=%
}}
\end{center}
\caption[]{ \footnotesize
 The $B$ dependence of the pp total cross section $\sigma_{pp}$. The
 five curves are lines  of constant  $\spai$,  of 440, 490, 540, 590 and
 640 mb---the central value is the {\em published} Fly's Eye value, and the others
 are $\pm 1\sigma$ and $\pm 2\sigma$. The dashed curve is a plot of our
 QCD-inspired fit of $B$ against $\sigma_{pp}$.  The dot is our fitted value for
 $\sqrt s=30$ TeV, the Fly's Eye energy.}
\label{fig:p-air}
\end{figure}

 The extraction of the pp cross section from the cosmic ray data is a two
 stage process. First, one calculates the $p$-air total cross section from
 the  inelastic cross section inferred in Eq.\,(\ref{eq:Lambda_m}),  where
\begin{equation}
\spai = \spa - \spae - \spaqe \,.  \label{eq:spa}
\end{equation}
 Next, the Glauber method\,\cite{yodh} transforms the
 value of $\spai$ into a proton-proton total cross section $\sigma_{pp}$;
 all the necessary steps are calculable in the theory, but depend sensitively  
on a knowledge of $B$, the slope of
 ${d\sigma_{pp}^{\rm el}\over dt}$, the  $pp$ differential elastic scattering  
cross section, where
\begin{equation}
B = \left[ {d\over dt} \left(\ln{d\sigma_{pp}^{\rm el}\over dt}\right)
 \right]_{t=0} \,.
\end{equation}
In Eq.\,(\ref{eq:spa})
  the cross section for particle production is supplemented with
$\spae$ and $\spaqe$,
 the elastic and quasi-elastic cross section, respectively, as calculated by the
 Glauber theory, to obtain the total cross section $\spa$.
We show in Fig.\,\ref{fig:p-air}%
plots of $B$ as a function of 
 $\sigma_{pp}$, for 5 curves of different values of $\spai$.
 This summarizes the reduction procedure
 from the measured quantity $\Lambda_m$ (of Eq.\,\ref{eq:Lambda_m}) to  
$\sigma_{pp}$\cite{gaisser}.
 Also plotted in Fig.\,\ref{fig:p-air} is a curve (dashed) of $B$ {\em vs.}
 $\sigma_{pp}$ which will be discussed later. 
 Two significant drawbacks of this extraction method are that one needs:
\begin{enumerate}
\item{
 a model of
 proton-air interactions to complete the loop between the measured
 attenuation length $\Lambda_m$ and the cross section $\spai$,
 {\em i.e.,} the value of $k$ in Eq. (\ref{eq:Lambda_m}).}
\item{a simultaneous relation between $B$ and  $\sigma_{pp}$ at very high  
energies---well above the region currently accessed by accelerators.}
\end{enumerate}
 A proposal
 to minimize the impact of theory on these needs is the topic of this note.

\begin{figure}[h] %Fig. 2
\begin{center}
\mbox{\epsfig{file=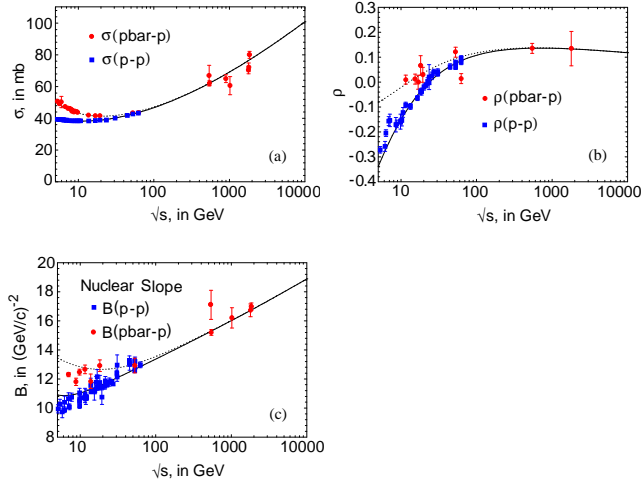%
            ,width=3.5in,bbllx=12pt,bblly=208pt,bburx=600pt,bbury=700pt,clip=%
}}
\end{center}
\caption[]{ \footnotesize
 The simultaneous QCD-inspired fit of total cross section
 $\sigma_{pp}$, $\rho$ and $B$ {\em vs.} $\sqrt s$, in GeV, for pp
 (squares) and $\bar {\rm   p}$p (circles) accelerator data:
 (a) $\sigma_{pp}$, in mb,\ \  (b) $\rho$,\ \  (c) Nuclear slope $B$,
 in GeV$^{-2}$}
\label{fig:ppcurves}
\end{figure}

We have constructed a QCD-inspired parameterization of the forward
 proton-proton and proton-antiproton scattering amplitudes\,\cite{block}
 which is analytic, unitary and fits all accelerator data\cite{orear} of  
$\sigma_{\rm tot}$, $B$
 and $\rho$, the ratio of the real-to-imaginary part of the forward
 scattering amplitude;  see Fig.\,\ref{fig:ppcurves}.%
In addition, the high energy cosmic ray data of Fly's Eye\,\cite{fly} and  
AGASSA\,\cite{akeno} experiments are also simultaneously used, {\em i.e.}, $k$  
from  Eq.~(\ref{eq:Lambda_m}) is also a fitted quantity---we refer to this fit  
as a {\em global} fit\,\cite{nonglobal}. We emphasize that in the global fit,  
all 4 quantities, $\sigma_{\rm tot}$, $B$,
 $\rho$ and $k$, are {\em simultaneously} fitted. Because our parameterization is
 both unitary and analytic, its high energy predictions are
 effectively model-independent, if you require that the proton is  
asymptotically a black disk. Using vector meson
 dominance and the additive quark models, we find further support for our QCD  
fit---it accommodates a wealth of
 data on photon-proton and photon-photon interactions without the
 introduction of new parameters\cite{eduardo}.  In particular, it also
 {\em simultaneously} fits $\sigma_{pp}$ and $B$, forcing a relationship
 between the two. Specifically, the $B$ {\em vs.} $\sigma_{pp}$ prediction
 of our fit completes the relation needed (using the Glauber model) between  
$\sigma_{pp}$  and $\spai$.
The percentage error in the prediction of $\sigma_{pp}$ at $\sqrt s=30$ TeV is  
$\approx 1.2$\%, due to the statistical error in the fitting parameters (see  
references \cite{block,eduardo}). A {\em major} difference between the present  
result, in which we simultaneously  fit the cosmic ray and accelerator data,  
and our earlier result\cite{nonglobal}, in which only accelerator data are  
used, is a {\em significant} reduction (about a factor of 2.5) in the errors of  
$\sigma_{pp}$ at $\sqrt s=30$.

\begin{figure}%Fig. 3
\begin{center}
\mbox{\epsfig{file=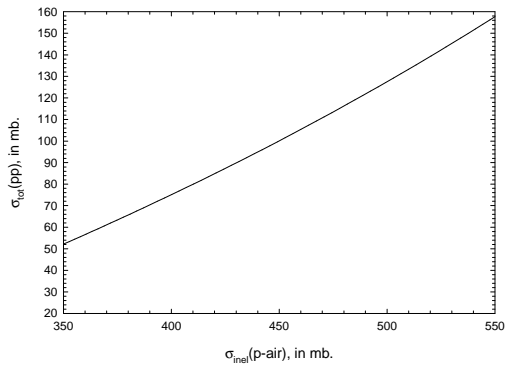%
              ,width=3in,bbllx=35pt,bblly=230pt,bburx=515pt,bbury=585pt,clip=%
}}
\end{center}
\caption[]{\footnotesize
A plot of the predicted total pp cross section $\sigma_{pp}$, in mb
 {\em vs.} the measured p-air cross section, $\spai$, in mb.
}
\label{fig:sigpp_p-air}
\end{figure}

In Fig.\,\ref{fig:sigpp_p-air},%
 we have plotted the values of
 $\sigma_{pp}$ {\em vs.} $\spai$ that are deduced from the
 intersections of our $B$-$\sigma_{pp}$ curve  with the $\spai$
 curves of Fig.\,1. Figure~\,\ref{fig:sigpp_p-air}
 allows the conversion of measured $\spai$ cross sections to $\sigma_{pp}$  
total cross sections. %
The percentage error in $\spai$ is $\approx 0.8$ \% near $\spai = 450 $mb, due  
to the errors
in $\sigma_{pp}$ and $B$ resulting from the errors in the fitting parameters.  
Again, the global fit gives an error of a factor of about 2.5 smaller than our  
earlier result\cite{nonglobal}, a {\em distinct} improvement.

 When we confront our predictions of the p-air cross sections ($\spai$) as a
 function of energy with published cross section measurements of the Fly's  
Eye\,\cite{fly} and AGASSA\,\cite{akeno} groups, %
%in Fig.\,\ref{fig:sigtodorpp} .%
%
%\begin{figure}%Fig. 4
%\begin{center}
%\mbox{\epsfig{file=sigtodorpp2.eps%
%              ,width=3in,bbllx=75pt,bblly=365pt,bburx=515pt,bbury=675pt,clip=%
%}}
%\end{center}
%\caption[]{\footnotesize
%\protect
%{  A plot of the QCD-inspired fit of the total nucleon-nucleon cross section
% $\sigma_{pp}$, in mb {\em vs.} $\sqrt s$, in GeV. The cosmic ray data that
% are shown have been converted from $\spai$ to $\sigma_{pp}$ using the
% results of Fig.~\ref{fig:sigpp_p-air} and the {\em published} values of $\spai$ .
%}
%}
%\label{fig:sigtodorpp}
%\end{figure}
%
%For inclusion in  Fig.\,\ref{fig:sigtodorpp}, we have  calculated
% the cosmic ray values of $\sigma_{pp}$ from the
% {\em published} experimental values of $\spai$, using the results
% of Fig.\,\ref{fig:sigpp_p-air}. We note the systematic underestimate
%
we find that the predictions   systematically are about one standard deviation  
below the {\em published} cosmic ray values.
It is at this point important to recall Eq.\,(\ref{eq:Lambda_m}) and remind  
ourselves that the measured experimental quantity is $\Lambda_m$ and {\em not}  
$\spai$.
 We emphasize that  the extraction of  $\spai$ from the
 measurement of $\Lambda_m$ requires {\em knowledge} of the parameter
 $k$. The measured depth $X_{\rm max}$ at which a shower reaches
 maximum development in the atmosphere, which is the basis of the
 cross section measurement in Ref.~\cite{fly}, is a combined measure
 of the depth of the first interaction, which is determined by
 the inelastic cross section, and of the subsequent shower development,
 which has to be corrected for. %
$X_{\rm max}$ increases logarithmically with energy with elongation
 rate ($\Delta X_{\rm max}$ per decade of Lab energy) of 50--60 g/cm$^2$
 in calculations with QCD-inspired hadronic interaction models. 
The position of $X_{\rm max}$
 directly affects the rate of shower attenuation with atmospheric depth,
 which is the alternative procedure for extracting $\spai$.
 The rate of shower development and its fluctuations
 are the origin of the deviation of $k$ from unity
 in Eq.\,(\ref{eq:Lambda_m}). Its predicted values range from 1.5 for a model
 where the inclusive cross section exhibits Feynman scaling, to 1.1
 for models with large scaling violations\,\cite{gaisser}. The comparison
 between prediction and experiment %in Fig.\,\ref{fig:sigtodorpp}
is further
 confused by the fact  that the AGASA\,\cite{akeno} and Fly's Eye\,\cite{fly}
 experiments used different values of $k$ in the analysis of their data,
 {\em i.e.,} AGASA used $k=1.5$ and Fly's Eye used $k=1.6$.

\begin{figure}%Fig. 5
\begin{center}
\mbox{\epsfig{file=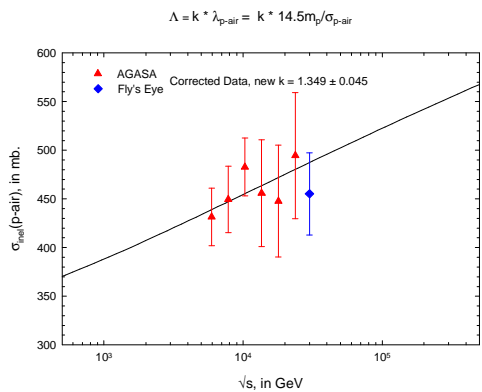%
              ,width=3in,bbllx=65pt,bblly=360pt,bburx=530pt,bbury=700pt,clip=%
}}
\end{center}
\caption[]{\footnotesize
\protect
{  The AGASA and Fly's Eye data for $\spai$, in mb,
 as a function of the energy, $\sqrt s$, in GeV, as found in our global fit,  
using the common value of $k=1.349$}
}
\label{fig:p-aircorrected2}
\end{figure}

 We therefore decided to let $k$ be a free parameter and to make a global fit  
to the accelerator and cosmic ray data, as emphasized earlier. This neglects  
the possibility
 that $k$ may show a weak energy dependence over the range measured. Recently,  
Pryke\cite{pryke} has made Monte Carlo model simulations that indicate that  
$k$ is compatible with being energy-independent. Using an energy-independent  
$k$, we find that $k=1.349\pm 0.045$, where the error in $k$ is the statistical  
error of the global fit.
By combining the results of Fig.\,\ref{fig:ppcurves}\,(a) and  
Fig.\,\ref{fig:sigpp_p-air}, we can predict the variation of $\spai$ with  
energy,  $\sqrt s$.
 In Fig.\,\ref{fig:p-aircorrected2}%
 we have {\em rescaled} the published high energy %\newpage \noindent
data for $\spai$ (using the common value of $k=1.349$), and plotted the  
revised data against our prediction of $\spai$ {\em vs.} $\sqrt s$.%
\begin{figure}[h]%Fig. 6
\begin{center}
\mbox{\epsfig{file=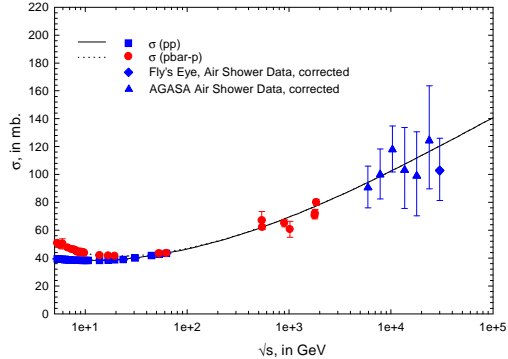%
              ,width=3in,bbllx=75pt,bblly=367pt,bburx=525pt,bbury=675pt,clip=%
}}
\end{center}
\caption[]{\footnotesize
\protect
{  A plot of the QCD-inspired fit of the total nucleon-nucleon cross section
 $\sigma_{pp}$, in mb {\em vs.} $\sqrt s$, in Gev. The cosmic ray data that
 are shown have been converted from $\spai$ to $\sigma_{pp}$ using the
 results of Fig.~\ref{fig:sigpp_p-air} and the common value of $k=1.349$,  
found from our global fit.
}
}
\label{fig:4sigtotcr}
\end{figure}

The plot of $\sigma_{pp}$ {\em vs.} $\sqrt s$, including the  rescaled cosmic  
ray data is shown in Fig.\,\ref{fig:4sigtotcr}. %
Clearly,
 we have an excellent fit, with good agreement between AGASA and Fly's Eye. In  
order to extract the cross sections' energy dependence from the cosmic ray  
data, the experimenters of course assigned energy values to their cross  
sections.  Since the cosmic ray spectra vary so rapidly with energy, we must  
allow for systematic errors in $k$ due to possible energy misassignments.  At  
the quoted experimental energy resolutions,
$\Delta{\rm Log}_{10}(E_{\rm lab}({\rm ev}))=0.12$ for AGASSA\cite{akeno} and
$\Delta{\rm Log}_{10}(E_{\rm lab}({\rm ev}))=0.4$ for Fly's Eye\cite{fly}, where  
$E_{\rm lab}$ is in electron volts,
we find from the curve in Fig.\,\ref{fig:p-aircorrected2} that $\Delta  
k/k=0.0084$ for AGASSA\cite{akeno} and $\Delta k/k=0.0279$ for Fly's  
Eye\cite{fly}. We estimate conservatively that experimental energy resolution  
introduces a systematic error in $k$ such that $\Delta k_{\rm systematic}=\sqrt  
{(\Delta k_{\rm AGASSA}^2+\Delta k_{\rm FLYSEYE}^2)/2}= 0.028$.  Thus, we  
write our final result as $k=1.349\pm 0.045\pm 0.028$, where the first error is  
statistical and the last error is systematic.

Recently, Pryke\cite{pryke} has published a comparative study of  high  
statistics simulated air showers for proton primaries, using four combinations  
of the MOCCA\cite{mocca} and
CORSIKA\cite{corsika} program frameworks, and SIBYLL\cite{sibyll} and  
QGSjet\cite{qgsjet} high energy hadronic interaction models. He finds $k=1.30  
\pm 0.04$ and $k=1.32 \pm 0.03$ for the CORSIKA-QGSjet and MOCCA-Internal  
models, respectively, which are in excellent agreement with our measured  
result, $k=1.349\pm 0.045\pm 0.028$.

Further, Pryke\cite{pryke} obtains  $k=1.15 \pm 0.03$ and $k=1.16 \pm 0.03$  
for the CORSIKA-SIBYLL and MOCCA-SIBYLL models, respectively, whereas the  
SYBILL\cite{gaisser} group finds $k=1.2$, which is not very different from the  
Pryke value.  However, the SYBILL-based models, with $k=$1.15--1.20, are  
significantly different from our measurement of $k=1.349\pm 0.045\pm 0.028$.
At first glance, this appears somewhat strange, since our
 model for forward scattering amplitudes and SIBYLL share the same
 underlying physics. The increase of the total cross section with
 energy to a black disk of soft partons is the shadow of increased
 particle production which is modeled in SYBILL by the production of (mini)-jets
 in QCD. The difference between the $k$ values of 1.15--1.20 and 1.349  
results from the very rapid rise of the $pp$ cross section in SIBYLL at the  
highest energies. This is an artifact of the fixed cutoff in transverse  
momentum used to compute the mini-jet production cross section, and is not a  
natural consequence of the physics in the model. There are ways to remedy this.

In conclusion, the overall agreement between the accelerator and the cosmic  
ray $pp$ cross sections with our QCD-inspired fit, as shown in  
Fig.\,\ref{fig:4sigtotcr}, is striking.
We find that the accelerator and cosmic ray $pp$ cross sections  are readily  
reconcilable using  a  value of $k=1.349\pm 0.045\pm 0.028$, which is both  
model independent and energy independent---this determination of $k$ severely  
constrains any model of high energy hadronic interactions. We predict high  
energy $\sigma_{pp}$ and $\spai$ cross sections that are accurate to $\approx$  
1.2\% and 0.8\%, respectively, at $\sqrt s=30$ TeV.

At the LHC ($\sqrt s=14$ TeV),
we predict $\sigma_{\rm tot}=107.9\pm 1.2$ mb for the total cross section,  
$B=19.59\pm 0.11$ (GeV/c)$^{-2}$ for the nuclear slope and $\rho=0.117\pm  
0.001$, where the quoted errors are due to the statistical errors
 of the fitting parameters.

In the near term, we look  forward
 to the possibility of repeating this analysis with the higher
 statistics of the HiRes\,\cite{HiRes} cosmic ray experiment that
 is currently in progress and the Auger\,\cite{Auger} Observatory.


\begin{thebibliography}{99} % Begin bibliography environment.
\frenchspacing

\bibitem{gaisser}
R. Engel {\it et al.}, Phys. Rev. D{\bf 58}, 014019, 1998.

\bibitem{yodh}
T. K. Gaisser {\it et al.}, Phys. Rev. D{\bf 36}, 1350, 1987.

\bibitem{block}
M. M. Block {\it et al.}, Phys. Rev. D{\bf 45}, 839, 1992.

\bibitem{orear}
We have now included in the accelerator data the new E-811 high energy cross  
section at the Tevatron:
C.~Avila {\it et al.}, Phys. Lett. B {\bf 445}, 419, 1999.
%
\bibitem{fly}
R. M. Baltrusaitis {\it et al.}, Phys. Rev. Lett. {\bf 52}, 1380, 1984.
%
\bibitem{akeno}
M. Honda {\it et al.}, Phys. Rev. Lett. {\bf70}, 525, 1993.
%
\bibitem{nonglobal}
In an earlier communication,  the accelerator data {\em alone} were fitted.   
Using the parameters from that fit, we
then made a {\em separate} fit of the cosmic ray data to the value of $k$; see  
M. M. Block {\it et al.}, e:-Print Archive:
{\bf hep-ph/9908222}, Phys. Rv. Lett. {\bf 83}, 4926, 1999. In this work, we  
make a {\em simultaneous} fit of the accelerator {\em and} the cosmic ray data,  
a much more complicated and very lengthy numerical analysis---but also a much  
superior physical analysis, resulting in greatly reduced errors in our  
predictions of high energy values of $\sigma_{pp}$ and $\spai$.
%
\bibitem{eduardo}
M. M. Block {\it et al.}, e-Print Archive: {\bf hep-ph/9809403},
 Phys. Rev. D{\bf 60}, 054024, 1999. %changed MMB 9/28/99
%
%\bibitem{eastop}
%M. Aglietta {\it et al}, Proc 25th ICRC (Durban)
%{\bf 6}, 37, 1997.
%
\bibitem{pryke}
C. L. Pryke, (2000), e-Print Archive: {\bf astro-ph/0003442}.
%
\bibitem{mocca} A. M. Hillas, Nuc. Phys. B (Proc. Suppl.) {\bf 52B}, 29, 1997. 
%
\bibitem{corsika}
J. Knapp {\em et al.}, Reports {\bf FZKA 6019} and {\bf FZKA 5828} (1998 and  
1996), Forschungszentrum Karlsruhe. Available from {\tt  
http://www-ik3.fzk.de/\~{ }heck/corsika/}.
%
\bibitem{sibyll}
R. S. Fletcher {\it et al.}, Phys. Rev. D{\bf 50}, 5710, 1994.
%
\bibitem{qgsjet}
N. N. Kalmykov {\em et al.}, Nuc. Phys. B (Proc. Suppl.) {\bf 52B}, 17, 1997.
\bibitem{HiRes}
See http://sunshine.chpc.utah.edu/research/cosmic/hires/
%
\bibitem{Auger}
The Pierre Auger Project Design Report, Fermilab report (Feb. 1997).

\end{thebibliography}
\end{document}